\documentclass[twocolumn,pra,showpacs,aps]{revtex4-1}

\usepackage[english]{babel}
\usepackage[ansinew]{inputenc}
\usepackage{times}
\usepackage{graphicx}
\usepackage{graphics}
\usepackage{amsmath}
\usepackage{amsfonts}
\usepackage{amssymb}
\usepackage{dsfont}
\usepackage{epstopdf}
\usepackage{makeidx}
\usepackage{subfigure}
\usepackage{color}
\usepackage{pgf}
\usepackage{bm}
\usepackage{tikz} 
\usepackage[normalem]{ulem}

\newcommand{\be}{\begin{equation}}
\newcommand{\ee}{\end{equation}}
\newcommand{\bea}{\begin{eqnarray}}
\newcommand{\eea}{\end{eqnarray}}

\makeindex
\begin{document}
\title{Molding the flow of light with a magnetic field: plasmonic cloaking and directional scattering}
\author{W. J. M.  Kort-Kamp}
\affiliation{Instituto de F\'{\i}sica, Universidade Federal do Rio de Janeiro,
Caixa Postal 68528, Rio de Janeiro 21941-972, RJ, Brazil}
\author{F. S. S. Rosa}
\affiliation{Instituto de F\'{\i}sica, Universidade Federal do Rio de Janeiro,
Caixa Postal 68528, Rio de Janeiro 21941-972, RJ, Brazil}
\author{F. A. Pinheiro}
\affiliation{Instituto de F\'{\i}sica, Universidade Federal do Rio de Janeiro,
Caixa Postal 68528, Rio de Janeiro 21941-972, RJ, Brazil}
\author{C. Farina}
\affiliation{Instituto de F\'{\i}sica, Universidade Federal do Rio de Janeiro,
Caixa Postal 68528, Rio de Janeiro 21941-972, RJ, Brazil}

\date{\today}

\begin{abstract}

We investigate electromagnetic scattering and plasmonic cloaking in a system composed by a dielectric cylinder coated with a magneto-optical shell. In the long-wavelength limit we demonstrate that the application of an external magnetic field can not only switch on and off the cloaking mechanism but also mitigate losses, as the absorption cross-section is shown to be minimal precisely at the cloaking operation frequency band.  We also show that the angular distribution of the scattered radiation can be effectively controlled by applying an external magnetic field, allowing for a swift change in the scattering pattern. By demonstrating that these results are feasible with realistic, existing magneto-optical materials, such as graphene epitaxially grown on SiC, we suggest that magnetic fields could be used as an effective, versatile external agent to tune plasmonic cloaks and to dynamically control electromagnetic scattering in an unprecedented way, we hope that these results may find use in disruptive photonic technologies.

\end{abstract}
\maketitle
\section{Introduction}
Classical electrodynamics and its applications have experienced notable progresses in the last decade following the introduction of metamaterials. Indeed metamaterials have permitted not only the discovery of novel physical phenomena but also the development of applications that allow an unprecedented control of electromagnetic (EM) waves, far beyond to what can be achieved with natural media~\cite{zheludev2012}. Among these applications, EM cloaking is arguably the most fascinating as the idea of rendering an object invisible has fueled human imagination for many years. Nowadays there are a number of approaches to cloaking, proving that a properly designed metamaterial can strongly suppress EM scattering by an object for any incidence angle so that it will be practically undetectable at a given frequency.

The coordinate-transformation method~\cite{pendry2006,leonhardt2006,schurig2006} and the scattering cancellation technique~\cite{alu2005,alu2008,edwards2009,filonov2012,chen2012,rainwater2012,KortKamp2013,nicorovici1994} were the first approaches that have successfully demonstrated that EM cloaking is possible. The coordinate-transformation method, introduced by the pioneer works of Pendry~\cite{schurig2006} and Leonhardt~\cite{leonhardt2006}, is grounded in the emerging field of transformation optics~\cite{chen2010}. It requires metamaterials with anisotropic and inhomogeneous profiles, which are able to bend the incoming EM radiation around a given region of space, rendering it invisible to an external observer. This method has been first experimentally realized for microwaves~\cite{schurig2006}, and later extended to infrared and visible frequencies~\cite{valentine2009,ergin2010}. The scattering cancellation technique, which constitutes the basis for the development of plasmonic cloaks, was proposed in~\cite{alu2005}. Applying this technique, a dielectric or conducting object can be effectively cloaked by covering it with a homogeneous and isotropic layer of plasmonic material with low-positive or negative electric permittivity. In these systems, the incident radiation induces a local polarization vector in the shell that is out-of-phase with respect to the local electric field so that the in-phase contribution given by the cloaked object may be partially or totally canceled~\cite{alu2005,alu2008,chen2012}. Experimental realizations of cylindrical plasmonic cloaks for microwaves exist in 2D~\cite{edwards2009} and 3D~\cite{rainwater2012}, paving the way for many applications in camouflaging, low-noise measurements, and non-invasive sensing~\cite{alu2008,chen2012}. Other approaches to EM cloaking for different frequency ranges exist, such as mantle cloaking~\cite{alu2009mantle,soric2013} and waveguide cloaking~\cite{tretyakov2009}, which have been both experimentally implemented.

Despite the success of the existing cloaking techniques, they generally suffer from practical physical limitations, namely the detrimental effect of losses and the limited operation frequency bandwidth. In the particular case of the scattering cancellation technique the operation bandwidth rely on the plasmonic properties of the shell; as a result, once the cover is designed and constructed, the cloaking mechanism is only operational at a single (narrow) frequency band. As practical applications would often require more flexibility in the design and in operational bandwidth, proposals of tunable cloaks have been developed~\cite{peining,zharova2012,milton2009}. One route to tunable cloaking involves the use of a graphene shell~\cite{chen2011,farhat2013}. Another possible implementation of tunable plasmonic cloaks is based on the nonlinearity of plasmonic shells~\cite{monticone2013,zharova2012}. However, the disadvantage of this scheme is that its the effectiveness depends on a given range of intensities for
the incident excitation.

In the present paper we investigate an alternative route to achieving tunable plasmonic cloaks, based on magneto-optical effects. We demonstrate that, by investigating EM scattering by a dielectric cylinder coated with a magneto-optically active shell in the long-wavelength limit, the application of an external magnetic field can not only switch on and off the cloaking mechanism~\cite{kortkampPRL} but also minimize electromagnetic absorption, one of the major limitations of existing plasmonic cloaks. Indeed we show that absorption cross-section can be significantly smaller in the presence of an external magnetic field precisely at the operation frequency band. In addition we prove, by calculating differential scattering cross-section, that the angular distribution of the scattered radiation can be effectively controlled by applying an external magnetic field, allowing for a swift change in the scattering pattern. We also discuss possible realistic implementations of magneto-optical shell, such as covers made of graphene. These results suggest that magneto-optical effects could be exploited in the design of tunable, versatile plasmonic cloaks, allowing for a precisely control of light scattering under the influence of an external magnetic field.


\section{MODEL AND ANALYTICAL RESULTS}

\begin{figure}[h!]
  \centering
  \includegraphics[scale=0.40]{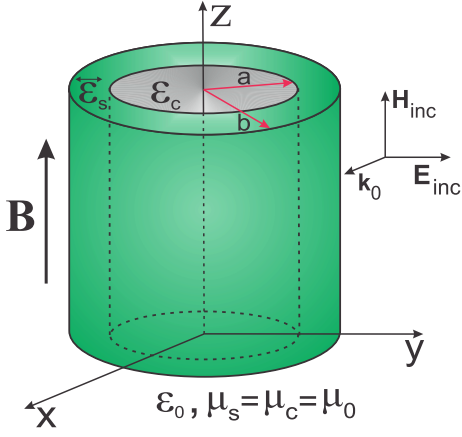}
  \caption{ The scattering system: an isotropic cylinder with dielectric constant $\varepsilon_c$ and radius $a$ coated by a magneto optical shell with permittivity tensor $\stackrel{\leftrightarrow}{\varepsilon}_s$ and outer radius $b>a$ under the influence of a static magnetic field ${\bf B}$ and a TM polarized monochromatic plane wave.}
  \label{Figura1}
\end{figure}

Let us consider an infinitely long, homogeneous cylinder with dielectric constant $\varepsilon_c$ and radius $a$ covered by a magneto optical shell with outer radius $b>a$. The symmetry axis of the cylinder coincides with the z-axis. Both the inner cylinder and the shell have a trivial permeability, meaning $\mu_c = \mu_s = \mu_0$. The whole system is subject to a uniform magnetic field ${\bf B}$ parallel to the z-axis, as depicted in fig. \ref{Figura1}. A monochromatic plane wave of angular frequency $\omega$ propagating in vacuum with its magnetic field parallel to the z-axis (TM polarization) impinges on the system normally to the main axis of the cylinder,
\begin{eqnarray}
&&{\bf H}_{\rm i} = H_{\rm i 0}\, e^{- i \omega (t - x/c)} \hat{z} \nonumber \\
&&{\bf E}_{\rm i} = - (\omega c) \, \hat{x} \times {\bf H}_{\rm i} = E_{\rm i 0}\,
e^{- i \omega (t - x/c)} \hat{y} ,
\end{eqnarray}

In the absence of the external magnetic field (${\bf B} = {\bf 0}$) the electromagnetic response of the shell to the impinging radiation can be well described by an isotropic permittivity tensor. However, when ${\bf B} \neq {\bf 0}$ the magneto-optical shell becomes anisotropic, and the shell's permittivity tensor acquires non-diagonal elements.  For the geometry shown in Fig. \ref{Figura1}, the permittivity tensor $\stackrel{\leftrightarrow}{\varepsilon}_s$ of the cloak can be cast in the form \cite{Stroud1990}
\begin{eqnarray}
\label{TensoresEeM}
\stackrel{\leftrightarrow}{\varepsilon}_s =
\left[
  \begin{array}{ccc}
    \varepsilon_{xx} & \varepsilon_{xy} & 0 \\
    \varepsilon_{yx} & \varepsilon_{yy} & 0 \\
    0 & 0 & \varepsilon_{zz} \\
  \end{array}
\right]
= \left[
  \begin{array}{ccc}
    \varepsilon_{s} & i\gamma_s & 0 \\
    -i\gamma_s & \varepsilon_{s} & 0 \\
    0 & 0 & \varepsilon_{zz}, \\
  \end{array}
\right]\, ,
\end{eqnarray}
where the specific expressions of $\varepsilon_s$, $\gamma_s$ and $\varepsilon_{zz}$ in terms of $\omega$ and $B = |{\bf B}|$ depend on the details of the materials used. At this point it is important to emphasize that the dependence of the dielectric tensor with $B$ plays a fundamental role in the scattering problem since the electromagnetic properties of the shell can be modified by controlling the strength and direction of the uniform magnetic field. In this way, the pattern of the scattered electromagnetic field by the system can be drastically altered in comparison to the case where ${\bf B}$ is absent.

In order to determine the scattered electromagnetic fields we have to solve the Maxwell equations,
\bea
&&\nabla \times {\bf E} = i\omega {\mu_0} {\bf H}\, \\ \cr
&&\nabla \times {\bf H}= -i\omega\!\! \stackrel{\leftrightarrow}{\varepsilon} \cdot {\bf E}\, ,
\eea
in the regions $\rho\leq a$, $a< \rho \leq b$, $b<\rho$ (hereafter denoted by regions 1, 2 and 3, respectively) with the appropriate boundary conditions at $\rho=a$ and $\rho=b$ \cite{BohrenHuffman}. Despite the anisotropic properties of the shell, for a TM normally incident plane wave the Maxwell equations can be decoupled in all three media \cite{Monzon}, and the only nonvanishing field components are $E_{\rho}(\rho,\varphi)$, $E_{\varphi}(\rho,\varphi)$ and $H_z(\rho,\varphi)$.
More explicitly, the z-component of the magnetic field in the i-th region, $H_z^{(i)}(\rho,\varphi)$, satisfies \cite{Monzon}
\begin{eqnarray}
&& \!\!\!\!\!\!(\stackrel{\leftrightarrow}{\varepsilon}_i)_{xx}  \dfrac{\partial^2H_z^{(i)}}{\partial x^2} +
(\stackrel{\leftrightarrow}{\varepsilon}_i)_{yy}\dfrac{\partial^2H_z^{(i)}}{\partial y^2} \   \cr\cr
&+&[(\stackrel{\leftrightarrow}{\varepsilon}_i)_{xy} + (\stackrel{\leftrightarrow}{\varepsilon}_i)_{yx}] \dfrac{\partial^2H_z^{(i)}}{\partial x \partial y}
+ \omega^2 \alpha_i{\mu}_0 H_z^{(i)}=0\, ,
\label{EquacaoCampoMagnetico}
\end{eqnarray}
with the remaining components of the electric field given by~\cite{Comment1}
\bea
\label{EquacaoCampoEletrico1}
E^{(i)}_{\rho}(\rho,\varphi) &=& \dfrac{i}{\omega\alpha_i}\left\{(\stackrel{\leftrightarrow}{\varepsilon}_i)_{xy}\dfrac{\partial H_z^{(i)}}{\partial \rho} +
\dfrac{(\stackrel{\leftrightarrow}{\varepsilon}_i)_{xx}}{\rho} \dfrac{\partial H_z^{(i)}}{\partial \varphi} \right\}\, , \\ \cr
E^{(i)}_{\varphi}(\rho,\varphi) &=& \dfrac{-i}{\omega\alpha_i}\left\{(\stackrel{\leftrightarrow}{\varepsilon}_i)_{xx} \dfrac{\partial H_z^{(i)}}{\partial \rho} -
\dfrac{(\stackrel{\leftrightarrow}{\varepsilon}_i)_{xy}}{\rho} \dfrac{\partial H_z^{(i)}}{\partial \varphi} \right\}\, ,
\label{EquacaoCampoEletrico2}
\eea
where $\stackrel{\leftrightarrow}{\varepsilon}_1 = \varepsilon_c \!\!\stackrel{\leftrightarrow}{1}$, $\stackrel{\leftrightarrow}{\varepsilon}_2 = \stackrel{\leftrightarrow}{\varepsilon}_s$, $\stackrel{\leftrightarrow}{\varepsilon}_3 = \varepsilon_0 \!\!\stackrel{\leftrightarrow}{1}$  and
\bea
\label{alpha}
\alpha_i := (\stackrel{\leftrightarrow}{\varepsilon}_i)_{xx}(\stackrel{\leftrightarrow}{\varepsilon}_i)_{yy}-
(\stackrel{\leftrightarrow}{\varepsilon}_i)_{xy}(\stackrel{\leftrightarrow}{\varepsilon}_i)_{yx}\, .
\eea

Given the anti-symmetry of $ \stackrel{\leftrightarrow}{\varepsilon}_s$, the term containing the cross derivative in (\ref{EquacaoCampoMagnetico}) vanishes even within the shell, and therefore $H_z^{(i)}$ satisfies a Helmholtz equation,
\begin{equation}
(\nabla^2 + k_i^2)H_z^{(p)} = 0\,\, ,
\label{Helmholtz}
\end{equation}
with the modulus of the wave vector given by
\begin{equation}
k_i = \omega \sqrt{\dfrac{{\mu}_0\alpha_i}{(\stackrel{\leftrightarrow}{\varepsilon}_i)_{xx}}}\, .
\label{wavevector}
\end{equation}

The solution of the Helmholtz equation in cylindrical
coordinates is well known and can be written as \cite{BohrenHuffman}
\begin{eqnarray}
H_z^{(1)}(\rho,\varphi) &=& \displaystyle{\sum_{m=-\infty}^{+\infty}} A_m J_m(k_c\rho)e^{im\varphi}\;\;\;\;  (\rho \leq a)\, ,
\label{CampoNoCilindroInterno}\\
H_z^{(2)}(\rho,\varphi) &=& \displaystyle{\sum_{m=-\infty}^{+\infty}} i^m\left\{B_m J_m(k_s\rho)\right.  \cr\cr
 &+& \left. C_m N_m(k_s\rho)\right\}e^{im\varphi}\;\;\;\;  (a<\rho<b)\, ,
\label{CampoNaCasca}\\
H_z^{(3)}(\rho,\varphi) &=& \displaystyle{\sum_{m=-\infty}^{+\infty}} i^m\left\{J_m(k_0\rho) \right. \cr\cr
&+& \left. D_m H_m^{(1)}(k_0\rho)\right\}e^{im\varphi}\;\;\;\; ( b \leq \rho)\, ,
\label{CampoEspalhado}
\end{eqnarray}
where $J_m(x)$, $N_m(x)$ and  $H_m^{(1)}(x)$ are $m$th-order cylindrical Bessel,
Neumann and Hankel (first kind) functions, respectively \cite{Abramowitz}, and
coefficients $A_m$, $B_m$, $C_m$ and $D_m$ are to be determined by
imposing the usual electromagnetic boundary conditions on the transverse
components of ${\bf E}$ and ${\bf H}$, namely,
\bea
&&\hspace{-10pt} H_z^{(1)}(a,\varphi) = H_z^{(2)}(a,\varphi) \;\; , \;\;\;
H_z^{(2)}(b,\varphi) = H_z^{(3)}(b,\varphi) \nonumber \\
&&\hspace{-10pt} E_{\varphi}^{(1)}(a,\varphi) = E_{\varphi}^{(2)}(a,\varphi)
\;\; , \;\;\;
E_{\varphi}^{(2)}(b,\varphi) = E_{\varphi}^{(3)}(b,\varphi) \, .
\label{CondicaoContorno1}
\eea
Our goal is to calculate the  differential scattering cross section per length $dQ_{sc}/d\varphi$ and the total scattering and extinction cross sections efficiencies $Q_{\textrm{sc}}$ and $Q_{\textrm{ext}}$, given by \cite{BohrenHuffman}
\begin{eqnarray}
\label{differential}
\dfrac{dQ_{sc}}{d\varphi} &=& \dfrac{2}{k_0\pi}\left|\sum_{-\infty}^{\infty}i^m D_m e^{-i\pi/4}e^{im(\varphi-\pi/2)}\right|^2\, ,\\
\label{QscQext}
Q_{\textrm{sc}} &=& \dfrac{2}{k_0b}\,\displaystyle{\sum_{-\infty}^{+\infty}|D_m|^2}\, , \\
\label{Qext}
Q_{\textrm{ext}} &=&
\dfrac{2}{k_0b}\,\displaystyle{\sum_{-\infty}^{+\infty}{\textrm{Re}}(D_m)} \, .
\end{eqnarray}

After long but straightforward calculations it is possible to show that
$D_m$ can be put in the form
\begin{equation}
D_m=\dfrac{U_m}{V_m}\, ,
\label{Dm}
\end{equation}
where
\bea
&&\!\!\!\!\!\!\!\! U_m = \cr\cr &\,& \!\!\!\!\!\!\!\!\!\! \left|
  \begin{array}{cccc}
    J_m(k_c a) &-J_m(k_s a) & -N_m(k_s a) & 0 \\
    \frac{k_c}{\varepsilon_c} J'(k_c a) & -{\cal{J}}_m(k_s a) & -{\cal{N}}_m(k_s a) &0 \\
    0 & J_m(k_s b) &N_m(k_s b) & J_m(k_0 b) \\
    0 & {\cal{J}}_m(k_s b) & {\cal{N}}_m(k_s a) & \frac{k_0}{\varepsilon_0}J'_m(k_0 b) \\
  \end{array}
  \right|
  \label{Um}
\eea
and
\bea
&&\!\!\!\!\!\! \!\! V_m= \cr\cr &\,& \!\!\!\!\!\!\!\!\!\! \left|
  \begin{array}{cccc}
    J_m(k_c a) &-J_m(k_s a) & -N_m(k_s a) & 0 \\
    \frac{k_c}{\varepsilon_c} J'(k_c a) & -{\cal{J}}_m(k_s a) & -{\cal{N}}_m(k_s a) &0 \\
    0 & J_m(k_s b) &N_m(k_s b) & -H_m^{(1)}(k_0 b) \\
    0 & {\cal{J}}_m(k_s b) & {\cal{N}}_m(k_s a) & -\frac{k_0}{\varepsilon_0}{H'}_m^{(1)}(k_0 b) \\
  \end{array}
\right|
\label{Vm}
\eea
with
\bea
\hskip -0.5cm {\cal{J}}_m(x) &=& \dfrac{1}{\varepsilon_s^2-\gamma_s^2}\left[\varepsilon_s k_s J'_m(x) + \dfrac{m\gamma_s k_s}{x} J_m(x)\right]\, , \\ \cr
\hskip -0.5cm {\cal{N}}_m(x) &=& \dfrac{1}{\varepsilon_s^2-\gamma_s^2}\left[\varepsilon_s k_s N'_m(x) + \dfrac{m\gamma_s k_s}{x} N_m(x)\right]\, ,
\eea
where the primes in $J'_m(x)$, $N'_m(x)$, ${H'}_m^{(1)}(x)$ denote
differentiation with respect to the argument.

\section{Results and Discussions}

In this section we apply the previous results for electromagnetic scattering from coated cylinders to a realistic physical system: an inner cylinder of polystyrene coated with graphene epitaxially grown on SiC. It has been recently shown that graphene epitaxially grown on SiC naturally exhibits nanoscale inhomogenities that lead to a pronounced Terahertz plasmonic resonance~\cite{Crassee2012}. The excellent plasmonic properties and strong magneto-optical activity makes graphene grown on SiC an excellent material platform to investigate the effects of an external magnetic field on plasmonic cloaking~\cite{kortkampPRL}.

It is straightforward to show that the effective electric permittivity of graphene grown on SiC can be well described by a Drude-Lorentz model given by
\begin{eqnarray}
\varepsilon_s(\omega, B) &=& \varepsilon_0+\dfrac{I\sigma_{xx}^{2D}(\omega,B)}{\omega (b-a)} \\ \cr
\gamma_s(\omega, B) &=& \dfrac{\sigma_{xy}^{2D}(\omega,B)}{\omega(b-a)},
\end{eqnarray}
where \cite{Crassee2012}
\begin{eqnarray}
\sigma_{xx}^{2D}(\omega,B) &=& 3.5\sigma_0+ \dfrac{\sigma_{+}^{2D}(\omega,B)+\sigma_{-}^{2D}(\omega,B)}{2}\\ \cr
\sigma_{xy}^{2D}(\omega,B) &=& \dfrac{\sigma_{+}^{2D}(\omega,B)-\sigma_{-}^{2D}(\omega,B)}{2i}.
\end{eqnarray}
with $\sigma_0=e^2/4\hbar$ being the universal conductivity and
\begin{eqnarray}
\sigma_{\pm}^{2D}(\omega,B) = \dfrac{2d}{\pi} \dfrac{i}{\omega \mp \omega_c-\omega_0^2/\omega+i\Gamma}.
\end{eqnarray}
In the following calculations we employ parameters directly fitted from experimental results, including material losses \cite{Crassee2012}:
\begin{eqnarray}
\hbar d/\sigma_0 = 0.52 \textrm{eV}\, , \ \ \omega_0 = 9.9\times10^{12} \textrm{rad/s}\, , \cr\cr
\Gamma = 18.3\times10^{12} \textrm{rad/s}\, , \ \omega_c = 3.2\times10^{12}B\textrm{(T)} \textrm{rad/s} \nonumber \, .
\end{eqnarray}
It is worth mentioning that in the experiment described in Ref.~\cite{Crassee2012} the magnetic field is applied perpendicularly to the sample of graphene epitaxially grown on SiC. Here, since we are considering effective material parameters, the values extracted from the experimental data provide reasonable estimates to be used in the following numerical calculations.

For the inner polystyrene cylinder the material parameters are \cite{Hough1980}:
\begin{equation}
\dfrac{\varepsilon_{c}(\omega)}{\varepsilon_0} = 1 + \dfrac{\omega_{p1}^2}{\omega_{r1}^2 - \omega^2 - i\Gamma_{1}\omega} +\dfrac{\omega_{p2}^2}{\omega_{r2}^2 - \omega^2 - i\Gamma_{2}\omega},
\end{equation}
where
\begin{eqnarray}
\omega_{p1} &=& 1.11\times10^{14} \textrm{rad/s}\, , \ \ \omega_{r1} = 5.54\times10^{14} \textrm{rad/s}\, , \cr\cr
\omega_{p2} &=& 1.96\times10^{16} \textrm{rad/s}\, , \ \ \omega_{r2} = 1.35\times10^{16} \textrm{rad/s}\, , \cr\cr
\Gamma_1 &=& \Gamma_2 = 0.1 \times 10^{12} \textrm{rad/s}\nonumber\, .
\end{eqnarray}
Here again we employ realistic parameters, including losses, and take into account that polystyrene is weakly dispersive \cite{Hough1980}.
We also emphasize that in the following calculations we restrict ourselves to the dipole approximation, {\it i.e.} $k_0b\ll 1\, , \ k_cb\ll 1\, , \ k_sb\ll1$. With this purpose we set $b=0.1\lambda$  and $a=0.6b$. We have verified that this condition guarantees the validity of the dipole approximation and that the dominant scattering coefficients are indeed $D_0$ and $D_{\pm 1}$.

\subsection{Tuning plasmonic cloaking and enhancing electromagnetic scattering with a magnetic field}

\begin{figure}[h!]
  \centering
  \includegraphics[scale=0.42]{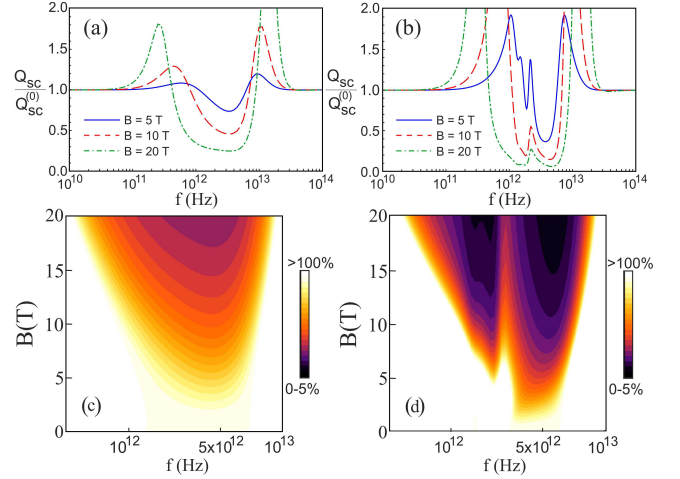}
  \caption{Scattering efficiency $Q_{sc}$ (normalized by its value in the absence of  {\bf B}, $Q_{sc}^{(0)}$) as a function of the frequency of the impinging wave for $B = 5$ T (solid line), $10$ T (dashed line) and $20$ T (dot-dashed line) for (a) the same losses than in graphene epitaxially growth on SiC ($\Gamma = 18.3\times 10^{12}$ rad/s), and (b) with losses ten times smaller losses ($\Gamma = 1.83\times 10^{12}$ rad/s). Contour plots of $Q_{sc}/Q_{sc}^{(0)}$ as a function of both frequency $f$ and magnetic field $B$  are shown in (c) and (d) for the same set of parameters chosen in (a) and (b), respectively.}\label{Figura2}
\end{figure}

{\bf }
In order to investigate magneto-optical effects on plasmonic cloaking, we show in Fig.~\ref{Figura2}a the scattering efficiency $Q_{sc}$ (Eq.~\ref{QscQext}) in the presence of an external magnetic field ${\bf B}$ (normalized to its value in the absence of ${\bf B}$, $Q^{(0)}_{sc}$) as a function of the frequency of the impinging wave for a polystyrene cylinder coated with graphene epitaxially grown on SiC; the material parameters correspond exactly to those extracted from the experiment~\cite{Crassee2012}. Figure~\ref{Figura2}a shows that the application of ${\bf B}$ drastically reduces $Q_{sc}$, increasing the plasmonic cloaking performance in comparison to the case without the magnetic field treated so far; this reduction can achieve 80\% for $B=20$ T, relative to the case where  $B=0$. For smaller magnetic fields this reduction is still impressive: approximately 55\% for $B=10$ T. The analysis of Fig.~\ref{Figura2}a also reveals that by increasing ${\bf B}$ it is possible to broaden the frequency band where cloaking occurs; for $B=20$ T this spectral band spans from 0.3 THz to 9.0 THz, which is particularly appealing for applications. Furthermore Fig.~\ref{Figura2}a shows that the application of ${\bf B}$ allows for a swift shift in the scattering pattern. Indeed, for a fixed frequency one can readily change, by applying ${\bf B}$, from a situation where cloaking occurs to one in which the system scatters considerably more radiation than in the case $B=0$. Figure~\ref{Figura2}c shows a contour plot of $Q_{sc}/Q^{(0)}_{sc}$ as a function of both $B$ and frequency in the terahertz range for the same system. Figure~\ref{Figura2}c confirms that the application of ${\bf B}$ can significantly reduce the scattering cross-section for a broad frequency band in the terahertz even for modest magnetic fields. Altogether, Figs.~\ref{Figura2}a and \ref{Figura2}c demonstrates that the application of an external magnetic field can not only reduce the scattering cross-section of a realistic system composed by existing magneto-optical materials (graphene) in a broad frequency range in the terahertz, but also can tune its scattering properties; these results suggest that this particular system could be employed in the design of a tunable magneto-optical cloaking device, first proposed in Ref.~\cite{kortkampPRL}.

With the purpose of analysing how to improve the tuning mechanism and the plasmonic cloaking performance, in Figs.~\ref{Figura2}b and \ref{Figura2}d we show $Q_{sc}/Q^{(0)}_{sc}$ for the same system but with $\Gamma = 1.83\times 10^{12}$ rad/s, {\it i.e.} with values of losses in the graphene layer ten times smaller than in the experiment of Ref.~\cite{Crassee2012}. From Figs.~\ref{Figura2}b and \ref{Figura2}d one can see that electromagnetic scattering can be almost totally suppressed in the presence of ${\bf B}$; this reduction attains up to 90\% for $B=15$ T in the frequency range from 1.2 THz to 5.6 THz (except for a frequency band of 0.6 THz centered at 2.5 THz). Even for moderate magnetic fields the scattering suppression is quite large, approximately 70\% (typical efficiency in plasmonic cloaking experiments \cite{edwards2009, rainwater2012}) for $B=9$ T in the frequency range from 1.4 THz to 5.6 THz (except for a frequency bandwidth of 0.4 THz centered at 2.4 THz). The effect of increasing the magnetic is twofold: to further reduce $Q_{sc}$ and to broaden the frequency band where this reduction occurs. In both cases reducing material losses contributes to enhance these effects.  Here again it is possible to drastically modify the scattering properties by varying the magnitude of the magnetic field, changing from a highly scattering situation to cloaking, whilst keeping the frequency constant.

\begin{figure}[h!]
  \centering
  \includegraphics[scale=0.42]{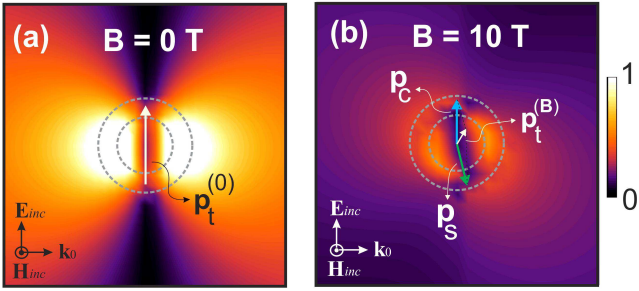}
  \caption{Spatial distribution of the scattered field $H_z$ in the $xy$ plane for (a) $B = 0$ T and (b) B = 20 T. The frequency of the incident wave is $f = 4$ THz. In panel (a) the vertical arrow indicates the total electric dipole ${\bf p}_t^{(0)}$ of the whole system for a vanishing magnetic field, whereas in panel (b) the arrows reprensent the electric dipole induced on the cylinder, ${\bf p}_c$, on the shell, ${\bf p}_s$, and the total electric dipole, ${\bf p}_t^{(B)}$ for $B \neq 0$.}

  \label{Figura3}
\end{figure}

Figure \ref{Figura3} shows the spatial distribution of the scattered field $H_z$ in the $xy$ plane. The losses are the same as those chosen in fig.~\ref{Figura2}b and the frequency is fixed at $f = 4$ THz. For this set of parameters, Fig.~\ref{Figura3}a shows that invisibility cannot be achieved for a vanishing magnetic field. In this case, the spatial distribution of $H_z$ is dipole-like since for ${\bf B} = {\bf 0}$ both the cylinder and the shell permittivities are isotropic and the total electric dipole ${\bf p}_t^{(0)}$  induced on the system is parallel to the impinging electric field, as shown in Fig.~\ref{Figura3}a. Figure \ref{Figura3}b shows that the presence of a magnetic field with magnitude $B = 10$ T strongly reduces the scattered field intensity for all observation angles, indicating that {\bf B} could play the role of an external agent to switch on and off the cloaking device operation. Besides, Fig. \ref{Figura3}b unveils the physical mechanism behind the scattering suppression: since polystyrene is an optically isotropic material, whose permittivity does not depend on {\bf B}, the dipole ${\bf p}_c$ induced on the cylinder is always parallel to the incident electric field; the shell, however, is optically anisotropic in the presence of an external magnetic field in such a way that the dipole ${\bf p}_s$ induced inside the shell is not parallel to the electric field anymore. The resulting electric dipole ${\bf p}_t^{(B)} = {\bf p}_c + {\bf p}_s$  of the whole system is therefore much smaller than ${\bf p}_t^{(0)}$ which explains why the scattered field is largely suppressed for $B \neq 0$. In addition, one can see that the scattering pattern in Fig.~\ref{Figura3}b is tilted in relation to the case shown in Fig~\ref{Figura3}a. The modification in the direction of maximum scattering is a direct consequence from the fact that ${\bf p}_t^{(B)}$ is not collinear with ${\bf E}_{inc}$ when ${\bf B}$ is applied, as it will be discussed in the next section.

\begin{figure}[h!]
  \centering
  \includegraphics[scale=0.42]{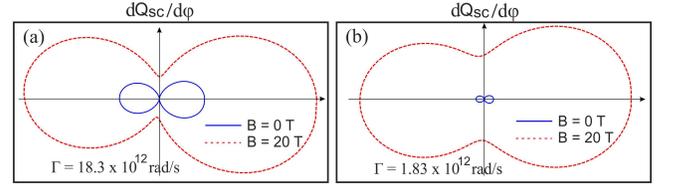}
  \caption{Differential scattering cross section for $B = 0$ T (solid line) and $B = 20$ T (dasehd line) for an incident wave of frequency $14$ THz and two different values of loss parameters: (a) $\Gamma = 18.3\times 10^{12}$ rad/s, and (b) $\Gamma = 1.83\times 10^{12}$ rad/s.}\label{Figura4}
\end{figure}

Figure \ref{Figura4} also shows the polar representation of the differential scattering cross-section given by Eq. (\ref{differential}) at frequency $14$ THz for two different values of the external magnetic field, namely, $B=0$ T and $B=20$ T. In panel (a) the dissipation parameter in the shell corresponds precisely to the one obtained from experimental data, $\Gamma = 18.3 \times 10^{12}$ rad/s, whereas in panel (b) it is 10 times smaller. For the chosen frequency here, the effect of ${\bf B}$ on the scattering properties of the system is quite different from the one previously discussed. Specifically, Figs.~\ref{Figura4}a and \ref{Figura4}b highlight the situation where EM scattering is strongly enhanced omnidirectionally when the magnetic field is present, regardless the values of $B$ and $\Gamma$. For $B = 20$ T and the set of parameters used in Fig.~\ref{Figura4}a the total scattering cross section efficiency can be 4 times larger than in the case where $B$ is absent, while for the same value of $B$ and smaller losses (Fig.~\ref{Figura4}b) the enhancement in the field intensities is almost 20 times greater. As discussed above, the effects of the external magnetic field on the scattering pattern are more pronounced for smaller losses. These results reinforce that magneto optical materials can be explored as a new and versatile platform for tuning the operation of plasmonic cloaking devices.

\subsection{Tailoring the scattering pattern with a magnetic field}

\begin{figure}[h!]
  \centering
  \includegraphics[scale=0.32]{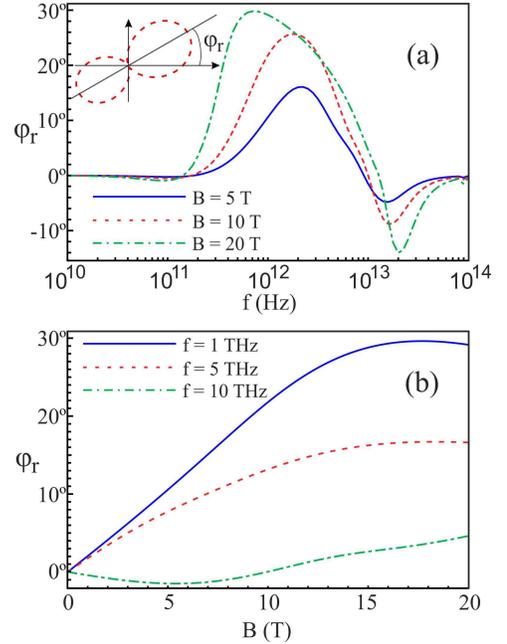}
  \caption{(a) Rotation angle $\varphi_r$ as a function of frequency for $B = 5$ T (solid line), $B = 10$ T (dashed line), and $B = 20$ T (dot-dashed line). The inset shows the definition of $\varphi_r$.  (b) The angle $\varphi_r$ as a function of the external magnetic field strength $B$ for three different frequencies of the impinging wave, $f = 1$ THz (solid line), $f = 5$ THz (dashed line) and $f = 10$ THz (dot-dashed line).}
  \label{Figura5}
\end{figure}

As shown in Fig.~\ref{Figura3}b the direction of the scattered radiation can be modified by the application of a magnetic field {\bf B}. The rotation of the scattered pattern can be quantified by the tilt angle $\varphi_r$ between the direction of maximum scattering and the horizontal axis (see the inset of Fig.~\ref{Figura5}a). In the case without losses and for nonmagnetic materials the coefficient $D_0$ vanishes and it follows from Eq. (\ref{differential}) that $\varphi_r$ is given by
\begin{equation}
\varphi_r = \dfrac{1}{2} \tan^{-1}\left[\dfrac{-\textrm{Im}(D_1D_{-1}^{*})}{\textrm{Re}(D_1D_{-1}^{*})}\right]\, .
\label{RotationAngle}
\end{equation}
Notice that $\varphi_r=0^o\, , \ 180^o$ when $B = 0$ T, as expected. Even if realistic losses are taken into account, we have numerically verified that for the set of materials we have chosen the above equation gives an excellent estimation to the rotation angle.

To investigate how an external magnetic field can modify and tailor the direction of the scattered radiation in magneto-optical systems we show in Fig.~\ref{Figura5}a the tilt angle $\varphi_{r}$ as a function of frequency $f$ for distinct values of ${\bf B}$. The parameters of the graphene layer are chosen from experimental data ~\cite{Crassee2012}.  It is evident from Figure~\ref{Figura5}a that the application of an external magnetic field induces a rotation of the scattering pattern. This rotation increases as one increases ${\bf B}$ and $\varphi_{r}$ exhibits a pronounced maximum in the terahertz frequency range. Increasing ${\bf B}$ has the effect of broadening the frequency band where the rotation occurs and of reducing the value of the position in frequency of the peak in $\varphi_{r}$; the maximal rotation can be as high as $\varphi_{r} \approx 30^{o}$ for $B=20$T and $f \approx 0.6$THz. It is worth mentioning that $\varphi_{r}$ changes its sign for higher frequencies ($f \gtrsim10$THz) and for all values of ${\bf B}$, as it can be seen from Fig.~\ref{Figura5}a. In Fig.~\ref{Figura5}b, $\varphi_{r}$ is calculated as a function of the strength of ${\bf B}$ for three distinct frequencies in the terahertz range. Figure~\ref{Figura5}b further demonstrates that radiation pattern can be significantly rotated by appliyng ${\bf B}$ for a broad frequency band in the terahertz range. In particular, the tilt angle $\phi_{r}$ increses from $0^{o}$ to $30^{o}$ as the magnetic field increases up to approximately 15 T for $f = 1$THz.

\begin{figure}[h!]
  \centering
  \includegraphics[scale=0.3]{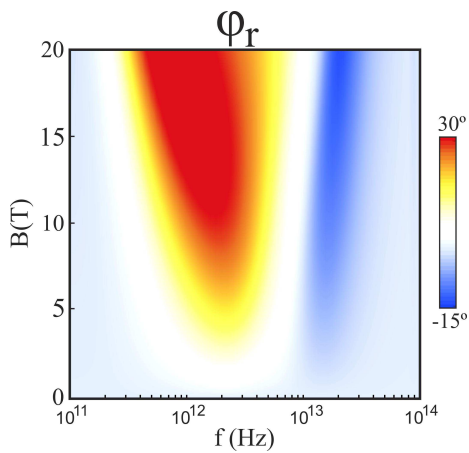}
  \caption{Rotation angle given by Eq. (\ref{RotationAngle}) as a function of both frequency $f$ and external magnetic field $B$ strenth.}
  \label{Figura6}
\end{figure}

In Fig.~\ref{Figura6} a contour plot of $\varphi_{r}$ as a function of both magnitude of ${\bf B}$ and frequency is shown. It is important to notice the large frequency band (from 1 THz to approximately 10 THz) where the rotation in the radiation pattern is maximal ($\varphi_{r} \approx 30^{o}$), which can be reached by applying a magnetic field of the order of 9 T. This frequency band tends to be broader with large magnetic fields. Even for modest magnetic fields (of the order of 5 T) a rotation of the order of $10^{o}$ can be achieved in the frequency range 1 THz to 4 THz.

\begin{figure}[h!]
  \centering
  \includegraphics[scale=0.42]{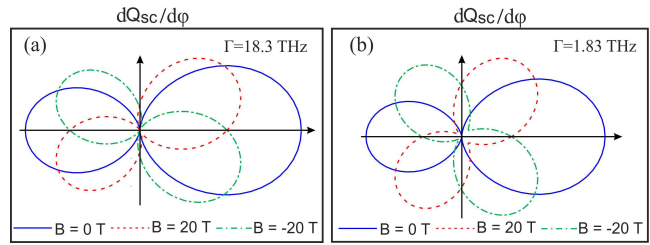}
  \caption{Differential scattering cross section for a incident wave of frequency $5$ THz and (a) $\Gamma = 18.3\times 10^{12}$ rad/s, and (b) $\Gamma = 1.83\times 10^{12}$. As the direction of the magnetic field is reversed whilst keeping constant its strength, the scattered pattern rotate up (a) $60^o$, and (b) $90^o$.}\label{Figura7}
\end{figure}

In Fig.~\ref{Figura7}a a polar representation of the scattering cross-section $dQ_{sc}/d\varphi$ is calculated for three different values of ${\bf B}$ and with the parameters of the graphene layer being taken from experimental data~\cite{Crassee2012}. The frequency of the incident electromagnetic plane wave is $5$ THz. Here one can appreciate the effects of reversing the direction of the magnetic field while keeping its magnitude constant: the radiation pattern drastically changes its direction, as $\varphi_{r}$ experiences a variation of 60$^o$ during this process. This effect of directioning the scattering pattern by applying ${\bf B}$ is even more pronounced if one reduces losses in the system, as demonstrated in Fig.~\ref{Figura6}b, where the dissipation parameter in the graphene layer is ten times smaller ($\Gamma = 1.83 \times 10^{12}$ rad/s). In this case $\varphi_{r}$ varies 90$^o$ as $B$ changes from 20 T to -20 T. Together with Figs.~\ref{Figura5} and \ref{Figura6}, these results demonstrate that one can achieve a high degree of tunability and control of the scattered radiation pattern, ultimately guiding the scattered light, by applying an external magnetic field in existing magneto-optical materials.

\subsection{Role of absorption}

\begin{figure}[h!]
  \centering
  \includegraphics[scale=0.42]{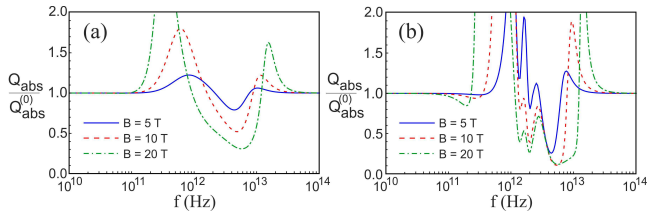}
  \caption{Absorption efficiency $Q_{abs}$ (normalized by its value in the absence of  {\bf B}, $Q_{abs}^{(0)}$) as a function of the frequency of the impinging wave for $B = 5$ T (solid line), $10$ T (dashed line) and $20$ T (dot-dashed line) for (a) the same losses than in graphene epitaxially growth on SiC ($\Gamma = 18.3\times 10^{12}$ rad/s), and (b) ten times smaller losses ($\Gamma = 1.83\times 10^{12}$ rad/s).}\label{Figura8}
\end{figure}

To investigate the effects of absorption on tunable cloaking with an external magnetic field in Fig.~\ref{Figura8} we calculate the absorption cross-section $Q_{abs}$ normalized by its value in the absence of ${\bf B}$, $Q^{(0)}_{abs}$, as a function of frequency for three values of the magnitude of ${\bf B}$. In Fig.~\ref{Figura8}a the material parameters of the graphene layer are exactly the same as in the experiments of Ref.~\cite{Crassee2012} whereas in Fig.~\ref{Figura8}b the loss parameter in such layer is ten time smaller. Contrasting Fig.~\ref{Figura8} and Fig.~\ref{Figura2} one can see that frequency regions where absorption and scattering minima occur approximately coincide when ${\bf B}$ is applied. This result is in contrast to what typically occurs in many cloaking devices: being based on resonant effects, cloaking mechanisms are usually associated with enhanced absorption, which degrades their performance. Figure~\ref{Figura8} indicates that absorption would not be a nuisance in this implementation of the magneto-optical tunable cloak. On the contrary, absorption is expected to be less significative precisely in the regions where cloaking occurs.

\section{Conclusion}

In summary, we have theoretically studied electromagnetic scattering and plasmonic cloaking in a system composed of a dielectric cylinder coated with a magneto-optical material. In particular, we consider a layer of graphene epitaxially grown on SiC, which has a pronounced terahertz plasmonic resonance due to morphological inhomogenities and exhibits strong magneto-optical activity~\cite{Crassee2012}. Using precisely the same experimental parameters of Ref.~\cite{Crassee2012}, including material losses, we demonstrate in the long-wavelength limit that tuning magneto-optical plasmonic cloaks with a external magnetic field, as first proposed in Ref.~\cite{kortkampPRL}, is indeed possible with realistic, existing magneto-optical materials. This tuning mechanism allows one not only to switch on and off cloaking by applying an external magnetic field but also to increase the frequency band where cloaking occurs. We demonstrate that cloaking typically shows up at frequency regions where the absorption cross-section is minimal. This result suggests that magneto-optical cloaks could circumvent one of the major problems of many cloaking devices, whose performance is typically decreased due to unavoidable material losses. We also show for this same system that it is possible to achieve a large enhancement of scattering by applying an external magnetic field, in addition to cloaking. The interchange between these two distinct scenarios (enhanced scattering and cloaking) can in turn be tuned by varying the magnetic field, allowing for a high degree of external control of the scattering properties. This control is also possible at the level of the angular distribution of scattered radiation, as the differential scattering radiation is shown to be highly tunable under the influence of an external magnetic field. As a result, the angular distribution of the scattered radiation can be significantly rotated by applying the magnetic field. Altogether, our findings may pave the way for the utilization of graphene in magneto-optical tunable cloaking devices, with versatile functionalities such as directional scattering, allowing for an unprecedented degree of control of the scattered electromagnetic radiation.

\section{Acknowledgments}

We thank T. J. Arruda, N. M. R. Peres, V. Barthem, and D. Givord for useful discussions, and FAPERJ, CNPq, and CAPES for financial support.



\begin{thebibliography}{99}
\bibitem{Comment1} Note that Eqs. (\ref{EquacaoCampoMagnetico}), (\ref{EquacaoCampoEletrico1}), (\ref{EquacaoCampoEletrico2}) and (\ref{alpha}) simplify considerably in the isotropic regions 1 and 3.
\end{thebibliography}
\end{document}